\documentclass[a4paper,14pt]{article}

\usepackage[T1]{fontenc}
\usepackage[utf8]{inputenc}
\usepackage{amsmath, amssymb, graphics}
\usepackage{extsizes} 
\title{\Large{Could Geoneutrinos Interact With the Geomagnetic Field?} }

\begin{document}

\author{C. A. B. Quintero$^{1}$ and J.A. Helayël Neto$^{2}$\\
\\
\normalsize{$^{1}$\emph{ Departamento de Geof\'isica, Observat\'orio Nacional (ON),}}\\
\normalsize{\emph{ Bairro Imperial de S\~ao Cristov\~ao, Rio de Janeiro, RJ, 20921-400, Brazil}}\\
\normalsize{$^{2}$\emph{ Centro Brasileiro de Pesquisas F\'isicas (CBPF),}}\\
\normalsize{\emph{ Urca, Rio de Janeiro, RJ, 22290-180, Brazil}}
\\
\normalsize{$^\ast$E-mail: bonilla@on.br, helayel@cbpf.br}}

\maketitle

\begin{abstract}
{In the present paper, we consider the possibility of interaction between geoneutrinos and the geomagnetic field, by adopting an approach based on the Dirac's equation with a non-minimal coupling that accounts for the magnetic interaction of the massive neutrinos. In our approach, we see that the magnetic interaction is controlled by a dimensionless parameter, $f\simeq 10^{-1}$, and we estimate the mean value of  this interaction to be of the order of $10^{-14}\ MeV^{2}$. } 
\end{abstract}

%***************************** SECTION INTRODUCTION
\section{Introduction}
Heat is the engine of the dynamics of the planet and  its magnetic field, so that it becomes impossible to split these two quantities \cite{Tolich:2012gua} \cite{bufet}. At present, Earth is releasing heat from its surface at a rate of about $47\ TW$ \cite{davies}  \cite{jaupart}. The principal contribution to the Earth loss are the secular cooling of the Earth and the decay of long-lived radioactive isotopes of uranium, thorium and potassium.\\
\indent The Bulk Silicate Earth (BSE) model establishes that the Earth's chemical composition must be the same as the one of meteorite chemical composition from primordial cloud. The chondritic ratio, $m_{Th}/m_{U}$, is one of the main characteristic of primordial cloud chemical composition and varies from  $2.6$ to $4.2$ \cite{lovering} \cite{kazunori}. BSE model gives values of mass for U, Th and K such that $m_{U}=0.81\times 10^{17}\ kg$, $m_{Th}=3.16\times 10^{17}\ kg$ and $m_{K}=0.49\times 10^{21}\ kg$. These amounts are distributed only in the Crust and Upper Mantel, this is, the BSE establishes that U and Th are not present in the core, although alternative hypothesis have been studied \cite{herdon} \cite{rama}. The main relevance of the chondritic ratio is that it is directly related to the flux of electronic antineutrinos, or geoneutrinos, from beta-decay of long-lived radioactive isotopes.\\ \indent
Geoneutrinos are detected via inverse beta-decay, $\bar{\nu_{e}}+p \longrightarrow e^{+} + n$, with an  energy threshold of $1.806\  MeV$. The prompt scintillation light from $e^{+}$ gives a measure of the $\bar{\nu_{e}}$-energy, $E_{\bar{\nu_{e}}}\simeq E_{p}+ \bar{E_{n}}+0.8MeV$, where $E_{p}$ is the prompt event energy including the positron kinetic and annihilation energies, and $\bar{E}$ is the average neutron recoil energy, $ \sigma \left(10 KeV\right)$  \cite{abe}. Actually, neutrinos are produced only in electron capture of $^{40}K$. In contrast to the Sun, Earth shines essentially in antineutrinos, ie, in geoneutrinos.\\ \indent
(Anti)neutrinos in the Standard Model (SM) do not have mass. As it is well-known, a massless, chiral neutrino cannot have a nonzero magnetic (and electric) dipole moment. However, beyond the SM, a massive Dirac or Majorana neutrino will, in general, exhibit magnetic moment \cite{fuji}. Therefore, the geoneutrino feels magnetic fields. Low \cite{low} \cite{barshay} has studied the interaction between particles through non-minimal couplings, opening up the possibility of understanding another kind of interaction, in this case, the spin of the neutrino with magnetic fields.\\ \indent
The purpose of this work is to study the possibility of an interaction of the massive geoneutrinos with the Geomagnetic field. For this purpose, we adopt a model from theoretical physics that is a relativistic generalization of the conventional spin-current, using non-minimal couplings. However, here, in our approach, we shall focus the discussion on the non-relativistic limit and then we will adapt our theoretical description for the case of geoneutrinos.\\ 
\indent This work is organized as follows. In Section 2, we describe the formalism and calculate the Lorentz force for geoneutrinos. In Section 3, we analyze the propagation of  geoneutrinos through the matter and we derive our approximateve expressions for the interaction between geoneutrinos and Geomagnetic field. Finally,our Final Considerations and Future Prospects are cast in Section 3.   
%*****************************| SECTION: PRESENTATION MODEL |******************************
%********************************************************************************************
\section{Presentation of the Model}
The interaction between the geoneutrinos and  matter is described by the weak nuclear force. geoneutrinos are electronic antineutrinos, hence,  they are fermions and their energy values are known. We are going to consider them as relativistic particles that obey the Dirac equation for spin-$\frac{1}{2}$ relativistic particles. In this approach, we are taking neutrinos as massive particles.   Here, we present, for the sake of completeness, the Dirac's equation for a Dirac or Majorana fermion with non-minimal couplings in addition. However, for our practical purposes in this paper, since we are concerned with neutrinos, we are going to finally set $e=0$, and focus on the parameters that govern the non-minimal couplings. Our investigation is not committed with the neutrino being a Majorana or a Dirac particle. In view of that,  the covariant Dirac equation with non-minimal couplings is given by,
 \begin{equation}
i \gamma^{\mu}\partial_{\mu}\psi-m\psi-eA_{\mu}\gamma^{\mu}\psi-fZ_{\mu}\gamma^{\mu}\gamma_{5}\psi+igZ_{\mu\nu}\Sigma^{\mu \nu}\gamma_{5}\psi=0.
\label{1}
\end{equation} 
Through this paper, we shall to work in natural unitis  ($\hbar=c=1$). Wherever necessary to recover the correct dimensions, we re-insert $\hbar$ and $c$ in the right places. The non-minimal couplings $f$ and $g$ are both nontrivial for either type of neutrino, once the neutrino field is described by Grassman value spinors. From the equation above, we get a Hamiltonian in which there appear new terms that correspond to a non-minimal interactions, 
\begin{eqnarray} \nonumber 
H\psi= & & \alpha^{i}\left[ -i\partial_{i}-eA^{i}-fZ^{i}\gamma_{5}\right]\psi + m\beta\psi+e\phi\psi \\ 
           & & +fZ_{0}\gamma_{5}\psi+ga^{i}\gamma^{i}\gamma_{5}\psi+2cgb^{i}\beta\gamma_{5}\Sigma^{i}\psi.
\label{2}
\end{eqnarray}
Before going on, it is necessary to set up some definitions for a better understanding of the parameters of the system. The field $\mathbf{Z_{\mu \nu}}$, present in the Hamiltonian, is a background field that can be split into $SO\left(3\right)$ irreducible representations. We define them as $\vec{a}$ such that $a_{i}$, and $\vec{b}$ such that $b_{k}$, thus:
\begin{eqnarray}\nonumber
a_{i}&=&-\partial_{t} Z_{i}-\partial_{i} Z_{0},\\
b_{k}&=&\epsilon_{ijk}\partial_{i}Z_{j}.
\end{eqnarray}
It is important to point out that $\vec{b}$ plays the rôle of a kind of background magnetic field that we decide to interpret as the geomagnetic field. 

%***************************************| SECTION: THE LORENTZ FORCE |**********************************
%*******************************************************************************************************
\subsection{The Lorentz-like Force}
We now consider the situation of geoneutrinos  in presence of the background fields, $\vec{a}$ and $\vec{b}$, that we shall relate to the wave geomagnetic field. After some algebraic manipulation, we arrive at the Lorentz force in terms of the external fields for the Hamiltonian of Eq(\ref{1}),
\begin{eqnarray}\nonumber
\mathfrak{F}^{i}&&= f\gamma_{5}\left(\partial_{i}Z_{0}-\partial_{t}Z^{i} \right)-f\Sigma^{j}\left( \partial_{j}Z^{i}-\partial_{j}Z^{i} \right) \\ \nonumber
% *******************************************************
&& +g\partial_{i}a^{j}\gamma^{j}\gamma_{5}\psi -2g\partial_{i}b^{j}\gamma^{j}-2ifmZ^{i}\gamma_{5}\beta \\ 
% *******************************************************
&& +4icgf Z^{i}b^{j}\beta\Sigma^{j}+2igfZ^{i}a^{j}\gamma^{j},
\label{lorentz}
\end{eqnarray}
where we do not see the conventional Lorentz force for the neutrinos chargeless (we have set $e=0$ in our Dirac's equation). On the other hand, we can see that there appears in the Lorentz force a new kind of interaction associated to the non-minimal couplings with the background fields $\vec{a}$ and $\vec{b}$. Here, the $\Sigma^{i}$'s are the well-known spin matrices which appear in a covariant Dirac's equation as the spatial components of $\Sigma^{\mu \nu}=\frac{i}{4}\left[\gamma^{\mu}, \gamma^{\nu}\right]$.\\
Since we are concerned with antineutrinos from beta-decay inside the Earth, we have to neglect all the terms in the Lorentz-like force related to the charge, and further on we shall focus on the terms that describe the interaction between the background magnetic field, that we are interpreting as a geomagnetic field, and the geoneutrinos, i.e., all the other terms are considered zero, 
\begin{equation}\nonumber
\mathfrak{F}^{i}= cf\gamma_{5}\left(\partial_{i}Z_{0}-\frac{1}{c}\partial_{t}Z^{i} \right) -cf\Sigma^{j}\left( \partial_{j}Z^{i}-\partial_{j}Z^{i} \right). 
\label{lorentz1}
\end{equation}
The second term $cf\Sigma^{i}\left( \partial_{i}Z^{j}-\partial_{j}Z^{i} \right) $, describes the interaction between the spin of the geoneutrino and the background magnetic field, $\vec{b}$. \\
Our next step is to work out its  value, for this purpose, we shall calculate the mean value of the term
\begin{equation}
cf \left\langle \Sigma^{i}b^{i}\right\rangle.
\label{magnetic}
\end{equation} 
There are many problems related to calculating the mean value of the Eq.(\ref{magnetic}). One of them is to know the value of the coupling constant $f$. Another important problem is to find out the most convenient wave function to describe geoneutrinos in this approach.
%****************************************| Section: The Brahama Aproach |******************************
%*************************************************************************************************
\subsection{Experimental data and estimation of parameters}
The wave function for a relativistic neutrino of mass $m$ and momentum $\vec{p}$ through vacuum can be written as \cite{bohem}
\begin{equation}
\nu\left( t\right)_{vacuum}=\psi_{0}e^{-it\left(\frac{m^{2}}{2p}\right)};
\label{wf_vacuum}
\end{equation}
however, when the geoneutrios propagate through matter, there is a change in the phase that  affects even its oscillation. In the approximation $\Delta m_{31}^{2}\sim \Delta m_{32}^{2}>>\Delta m_{21}^{2}$, the survival probability $P_{ee}$ for a electronic antineutrino in the vacuum is given by \cite{bohem}:
\begin{equation}
P_{ee}=\sin^{4}\theta_{13}+\cos^{4}\theta_{13}\left[1-\sin^{2}2\theta_{12}\sin^{2}\left(\frac{1.267 \Delta m_{21}^{2}L}{4E}\right)\right],
\end{equation}
where $\theta$ is the angle between the mass eigenstates and the weak eigenstates and $\Delta m^{2}$ is the difference of the squared mass eigenvalues. This equation is valid in the 2-flavor approximation. Taking that the diameter of the Earth is about $13000\ km$, for antineutrinos with energy $\sim 3\ MeV$, the oscillation length,
\begin{equation}
L_{0}\sim \pi c \hbar \frac{4E}{\Delta m^{2}_{21}},
\end{equation}
is of the order of $\sim 100\ km$, which is very small compared to Earth's diameter, and the effect of the neutrino oscillation to the total neutrino flux is well-averaged, given an overall survival probability of $P_{ee}\sim 0.54$ \cite{zav} \cite{fogli}.\\
\indent
Matter is composed by quarks and electrons. The electron neutrino has a especial behavior; it interacts with the electron via the exchange of the charged boson $W^{+}$. Neutrino-neutrino interactions constitute neutral currents and only take place with $Z^{0}$-exchange. The fact that electronic neutrino interacts in a different form has important consequences; one of them is that it feels a potential due to electrons and nucleons. Thus, the time-dependent Hamiltonian for a neutrino propagating through matter gets an extra term that modifies its phase and its related wave function as below:
\begin{equation}
\nu\left(t \right)_{matter}=\psi_{0}e^{-it\left(\frac{m^{2}}{2p}+\sqrt{2}G_{F}N_{e} \right)},
\end{equation}
where $G_{F}$ is the Fermi constant, $N_{e}=\frac{\rho N_{0}Z}{A}$, $\frac{Z}{A}$ is the average charge to mass ratio of the electrically  neutral matter and $N_{0}$ is the Avogadro number \cite{bohem}. Since the electron density in the Earth is not constant and  it shows moreover drastic changes in correspondence with boundaries of different Earth's layers, the behavior of the survival probability is not trivial and the equations have to be solved by numerical tracing. It has been set in \cite{enomoto} \cite{honda} that this so-called matter effect contribution to the average survival probability corresponds to an increase of about $2\%$ and the spectral distorsion is below $1\%$. Thus, the effect of the flavor oscillation on the total flux of geoneutrinos during propagation is $\sim 0.55$ with a very small spectral distortion, completely negligible for the precision of the current geoneutrino experiments.\\ 
\indent Notwithstanding  the description above, we are interested in setting up a scenario that includes the geomagnetic field. We are going to assume that the magnetic field is much smaller than $m^{2}_{p}/c$, where $m_{p}$ is the proton mass, with typical values $\simeq 10^{-13}\ MeV^{2}$, that is, the mean value of the geomagnetic field at Rio de Janeiro, Brazil, for example. As it has been mentioned above, we wish to point out the rôle of the geomagnetic field and its interaction through non-minimal coupling with antineutrinos from the inverse beta-decay. For this purpose, we are going to use the formalism developed in \cite{brahma1}.\\
There is an important reason to adopt this approach: when we analyze the energy spectrum, we can immediately see that the typical values of the geomagnetic field are very tiny, so that it is reasonable to consider that the effect of the geomagnetic field on geoneutrinos is perturbative,
\begin{equation}
E^{2}=m^{2}+p_{z}^{2}+2ne\mathfrak{B}^{z},
\label{bra2}
\end{equation}  
with $m$ the electron mass, and $p_{z}$ the electron momentum in the $z-$direction and $\mathfrak{B}^{z} \sim 10^{-13}MeV^{2}$ stands for magnetic field, as presented in \cite{brahma1}. In this approach, the magnetic field is considered along the $z$- direction. The maximum value of the parameter $n$ is
\begin{equation}
n_{max}=int\left\{ \frac{1}{2e\mathfrak{B}}\left(\left[\left(m_{n}-m_{p}\right)+E_{\bar{\nu}} \right]^{2}-m^{2}\right)\right\}\simeq 3\times 10^{16},
\label{bra3}
\end{equation} 
where $m_{p}$ is the proton mass, $m_{n}$ is the neutron mass, $E_{\bar{\nu}}$ is the energy of the antineutrino. Since $n$ can take  many values, we are going to consider, in this work, only the maximum value, since in some cases the minimum value  is zero. Thus, the electron moment in the $z$-direction is \cite{brahma1},
\begin{equation}
p_{z}=\sqrt{\left[\left(M_{n}-M_{p}\right)+E_{\bar{\nu}}\right]^{2}-m^{2}-2ne\mathfrak{B}}.
\label{bra4}
\end{equation} 
Replacing the values in the  equation Eq.(\ref{bra4}) above, we obtain $p_{z}\simeq 11.277\ MeV  $. 
Now, we can calculate, from Eq.(\ref{bra2}), the energy value $E_{max}\simeq 15.9637 MeV$. 
We calculate the currents $J_{i}$ and we get the analytic form given below:
\begin{eqnarray}\nonumber
J_{x} &= &0 \\ \nonumber
J_{y} &= &0\\ \nonumber
J_{z} &= & \frac{2\left(e\mathfrak{B}^{z}\right)^{\frac{3}{2}}}{E_{n}+m}p_{z},
\label{bra7}
\end{eqnarray}
with $J_{z\ max}\simeq 2\times 10^{-22} MeV$. Finally, we have calculated a set of parameters that will help us to calculate the mean value for geomagnetic field. In this formalism, the analytic expression for the mean value of the geomagnetic field projection is,
\begin{eqnarray}\nonumber
f<\Sigma^{x}\mathfrak{B}^{x}>_{x} &= &0 \\ \nonumber
f<\Sigma^{y}\mathfrak{B}^{y}>_{y} &= &0\\ 
f<\Sigma^{z}\mathfrak{B}^{z}>_{z} &= & f\mathfrak{B}^{z}\left[ 1+\frac{1}{\left(E_{n}+m\right)^{2}}p_{z}^{2}-\frac{2ne\mathfrak{B}^{z}}{\left(E_{n}+m\right)^{2}}\right].
\label{bra9}
\end{eqnarray}
Replacing the values in the  expression above, we obtain for the mean value the maximum and minimum values, namely:
\begin{eqnarray}\nonumber
f<\Sigma^{z}\mathfrak{B}^{z}>_{z\ max} &= &4.55921\times 10^{-14}\ MeV^{2}, \nonumber \\
f<\Sigma^{z}\mathfrak{B}^{z}>_{z\ min} &= & 1.26148\times 10^{-14}\ MeV^{2};
\label{bra10}
\end{eqnarray}
hence, in the spirit of the perturbative approach, we estimate the value of the parameter to be $f\simeq 10^{-1}$.\\
There are three classes of bulk silicate earth (BSE) models: the cosmochemical, geochemical and geodynamical \cite{Dye}. These models provide estimations for the global geoneutrino signal from the crust and mantle for $^{238}U$ and $^{232}Th$. The maximal value of the signal in the above description is $35\ TNU$. We follow Fiorentini \cite{Fiorentini} and estimate the signal in this approach arrive to $S_{T}\simeq 0.66\times 10^{48}\ TNU$ that is a unreal value for the signal.   

% ******************************|| SECTION: FINAL CONSIDERATIONS ||***************************************
% *****************************************************************************************************
\section{Final Considerations and Future Prospects}
A formulation based on non-minimal couplings allows us to explore the possibility of an interaction between geoneutrinos and the geomagnetic field. Attempts to understand how neutrinos interacts in Nature are pursued in \cite{modern1} \cite{modern2} \cite{modern3} \cite{max}. We investigate one face of the problem: their possible interaction with the geomagnetic field. In this context, we take an approach that highlights this aspect, and we place the interaction with the matter in a second plane, once it is related to oscillation length. But, immediately, there arise a question: is it possible that the combined action of the matter and the geomagnetic field has a more important rôle in the description of this problem? It is known now a days, that only oscillations between two families is considered in geoneutrinos; this is the same scenario when we consider the effects of the matter and the geomagnetic field together, even with a low value of the geomagnetic field. This is, the consideration about the two families still kept allow us to go deeper in this issue?\\
\indent In the scenario that we have proposed, we can see that the interaction is supported by a coupling parameter estimated to be of order $f\simeq10^{-1}$, and we conclude that the mean value of the magnetic interaction is $\sim 10^{-22}\ MeV^{2}$, a very small value. It is clear that, for to study this kind of interaction, in this approach, we have to consider the possibility of obtaining an enhanced flux of geoneutrinos in zones where the magnetic field is more intense, ie, the geomagnetic poles. Possibly, a phenomenon like this would have been detected already by the neutrinos detectors; but this is not happening.\\
Measurement of geoneutrino obtained from $1353$ days at Laboratory Nazionali del Gran Sasso (LNGS), in Italia, reports a signal of $38.8 \pm 12.0\ TNU$ with just a $6 \times 10^{-6}$ probability for a null geoneutrino measurement \cite{belini}. Different analyses using cosmochemical, geochemical and geodynamical approach seem to agree with the above value in the context of the BSE model. This model is based on the supposition of the a chondritic Earth, we could ask us  is this measurement support only  the consideration of a chondritic Earth with a ratio $m_{Th}/m_{U}\sim 4$ \cite{berzukov} \cite{rusov} \cite{meier}. We obtain a value for the signal of the geoneutrinos very high, an unrealistic value. We interpret that the overestimation of the geoneutrino signal value is due to the fact that the model only uses the geomagnetic field value for the calculus of the current and the flux of the geoneutrinos and not use any other geophysical parameter.\\
\indent It is an issue to be pursued  is to consider the specific case  of massive Majorana neutrinos, the family oscillations and the magnetism (actually, the magnetic dipole moment) of massive neutrinos. It would be interesting, as a follow-up of this paper, to discuss those matters in a geomagnetism scenario. These are open problems that might be the object of future investigations.

{}

% *******************************|| THE END ||************************************************************
% ******************************************************************************************************

\end{document}